\documentclass{article}
\usepackage[T1]{fontenc} 
\usepackage[utf8]{inputenc} 
\usepackage{ismir,amsmath,cite,url}
\usepackage{graphicx}

\usepackage{color}

\usepackage[firstpage]{draftwatermark}
\definecolor{lightgray}{rgb}{0.9,0.9,0.9}
\definecolor{darkgray}{rgb}{0.4,0.4,0.4}
\SetWatermarkFontSize{12pt}
\SetWatermarkScale{1.1}
\SetWatermarkAngle{90}
\SetWatermarkHorCenter{202mm}
\SetWatermarkVerCenter{170mm}
\SetWatermarkColor{darkgray}
\SetWatermarkText{Late-Breaking / Demo Session Extended Abstract, ISMIR 2024 Conference}

\usepackage{makecell}
\usepackage{multirow}

\def\lbd{}	

\title{A Mamba-Based Model for Automatic Chord Recognition}

\twoauthors
  {Chunyu Yuan} {CUNY Graduate Center \\ Computer Science \\}
  {Johanna Devaney} {Brooklyn College and CUNY Graduate Center  \\ Data Analysis \& Visualization and Music \\ {\tt johanna.devaney@brooklyn.cuny.edu}} 

\def\authorname{C. Yuan and J. Devaney}

\begin{document}

\maketitle
\begin{abstract}
In this work, we propose a new efficient solution, which is a Mamba-based model named BMACE (Bidirectional Mamba-based network, for Automatic Chord Estimation), which utilizes selective structured state-space models in a bidirectional Mamba layer to effectively model temporal dependencies. Our model achieves high prediction performance comparable to state-of-the-art models, with the advantage of requiring  fewer parameters and lower computational resources. 
\end{abstract}
\section{Introduction and Background}\label{sec:introduction}

Automatic chord recognition/estimation (ACE) has a long history in music information retrieval (MIR) research \cite{PauwelsOGS19_20YearsChords_ISMIR}. While the use of modern deep-learning techniques led to major improvements \cite{korzeniowski2016fully}, even the recent state-of-art approaches still experience a performance ceiling {\cite{McFeeB17_StructTrainingChords_ISMIR, park_bi-directional_2019, ito_harmonic_2021}. Some challenges in ACE that have been previously identified are the large number of label permutations \cite{McFeeB17_StructTrainingChords_ISMIR} and disagreements between expert annotators \cite{HumphreyB15_ChordInsights_ISMIR,condit18,koops19,koops17}, which is particularly true for rare chords \cite{NiMSD13_ChordSubjectivity_IEEE-TASLP,koops19}. While transformer-based models (e.g., \cite{park_bi-directional_2019} excel in capturing the necessary temporal dependencies for the ACE task, they also introduce significant computational overhead due to their quadratic complexity with respect to input length.
The increased complexity of transformer architectures, combined with their high memory and processing requirements, limits their usability in low-latency environments, such as real-time music analysis systems or embedded devices. Such applications call for a careful balance between model accuracy and computational efficiency.

In this paper, we evaluate how much improvement and compactness can be achieved on the ACE task by updating the model architecture, specifically by adding selective structured state-space models in a bidirectional Mamba layer. Specifically, we introduce BMACE (Bidirectional Mamba-based Network for Automatic Chord Estimation), a novel Mamba-based model that incorporates selective structured state-space models within a bidirectional Mamba layer to enhance the modeling of temporal dependencies. Notably, this model achieves performance comparable to its predecessors while utilizing fewer parameters, and lower computational costs. 

\begin{figure*}[h!]
\centering
\includegraphics[width=.8\textwidth] 
{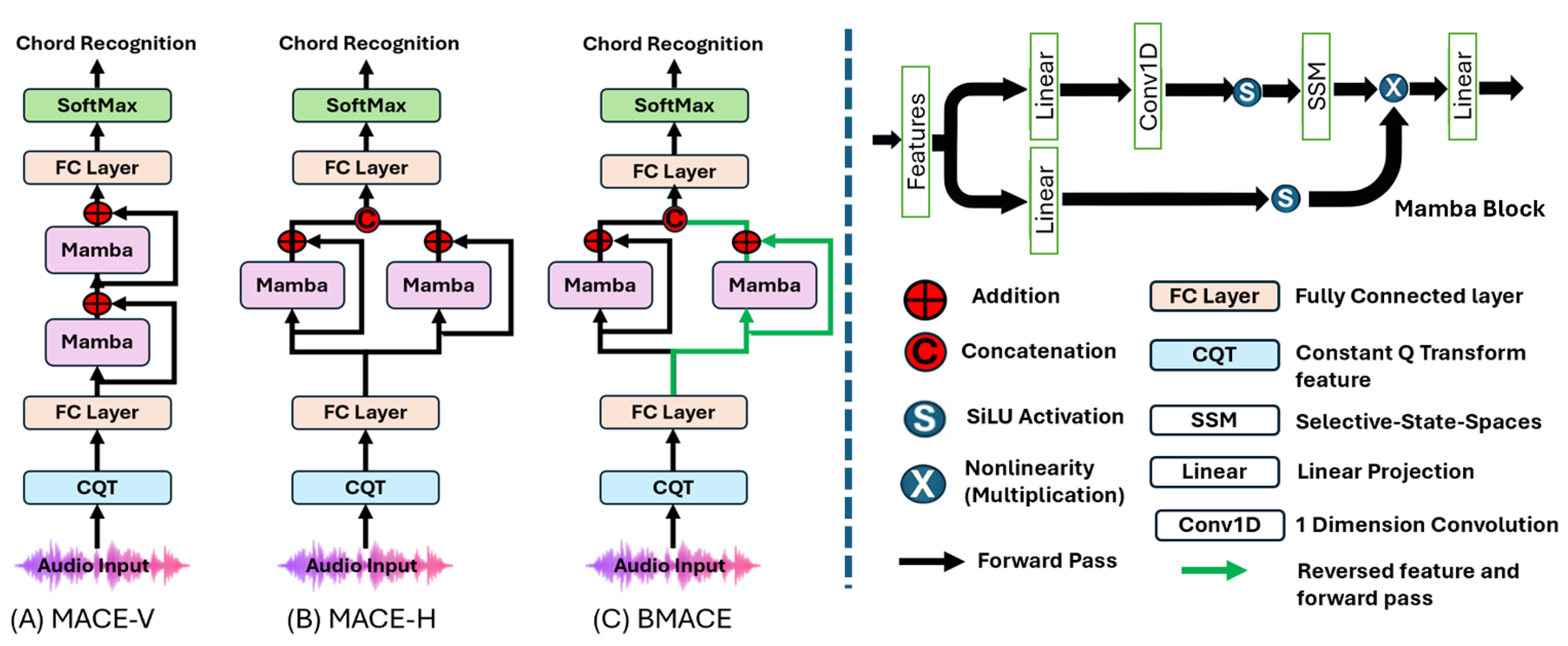} 
\caption{Architecture of Chord Recognition Models with Mamba Block. The diagram illustrates three variants, MACE-V (A), MACE-H (B), and BMACE (C). Each utilizes the Mamba block for improved feature processing. The Mamba block employs selective-state-spaces (SSM), SiLU activation, and 1D convolutions (Conv1D) for feature transformation. The figure highlights forward pass operations, feature reversal, addition, and concatenation mechanisms in the respective models, with fully connected (FC) layers leading to SoftMax output for chord recognition.}
\label{fig:variants}
\end{figure*}

\begin{table*}[h!]
 \begin{center}
 \footnotesize \small
 \renewcommand{\arraystretch}{1.5}
  \setlength{\tabcolsep}{2pt}
\begin{tabular}{c|cccc|ccccccccc}
\Xhline{3\arrayrulewidth}
\hline
\multirow{2}{*}{Model} & \multicolumn{4}{c|}{maj-min label type} & \multicolumn{9}{c}{large vocabulary label type}\\ \cline{2-14}
& Root↑  & Maj-min↑    &GFlops↓&Params↓& Root↑   & Thirds↑  & Triads↑  & Sevenths↑  & Tetrads↑  & Maj-min↑  & MIREX↑  &GFlops↓ &Params↓ \\ \Xhline{3\arrayrulewidth}

CRNN \cite{McFeeB17_StructTrainingChords_ISMIR} & 0.8185& \textbf{0.7796}& 0.0957 & 435,609 & 0.8026  & 0.7459 & 0.6384 & 0.6426 & 0.5448 & \textbf{0.7544} &0.7543&0.1038& 472,874\\ 
BTC \cite{park_bi-directional_2019} & 0.8202 & 0.7628 & 0.6282 &2,910,361& \textbf{0.8051} & \textbf{0.7524} & \textbf{0.6469} & 0.6506 & \textbf{0.5604} & 0.7531&0.7486& 0.6362 & 2,929,066\\
MACE-V & 0.7920& 0.7309 &  \textbf{0.0247} & \textbf{111,161} & 0.7739 & 0.715 & 0.6084 & 0.6137 & 0.5242& 0.7166&0.7057&\textbf{0.0328} &\textbf{129,866}\\
MACE-H &0.7898  & 0.7347 &  0.0261 & 114,361& 0.7833& 0.7211 &0.6188  & 0.6236 & 0.5314& 0.7304&0.7238&0.0422 & 151,626\\
BMACE & \textbf{0.8212} &0.7678  &  0.0261 & 114,361 & 0.8043 & 0.7455 & 0.6426 & \textbf{0.6526} &0.5571 & 0.7536&\textbf{0.7595}& 0.0422 &151,626\\
\hline
\end{tabular}
\end{center}
 \caption{Weighted Chord Symbol Recall (WCSR) scores for the performance of the CRNN \cite{McFeeB17_StructTrainingChords_ISMIR}, BTC \cite{park_bi-directional_2019}, and our three Mamba variants (MACE-V, MACE-H, and BMACE) on the uspop2002 dataset.}
 \label{tab:uspop} 
\end{table*}

\section{Bi-directional Mamba Network}\label{sec:mamba}

Inspired by the bidirectional Transformer, we propose a lightweight bidirectional Mamba-based network specifically designed for chord estimation/recognition: BMACE (Bidirectional Mamba-based network for Automatic Chord Estimation). The Mamba architecture was first introduced in late 2023 \cite{gu2023mamba} and has been gaining rapid momentum since its release. It has been applied to some speech \cite{jiang2024dual, li2024spmamba, quan2024multichannel} and some MIR \cite{bai2024two, chen2024musicmamba} tasks, but not yet for ACE. Mamba distinguishes itself from other models by eschewing the usual attention and MLP blocks for a more streamlined approach. This results in a model that is not only lighter and faster but also uniquely capable of scaling linearly with sequence length, an achievement that sets it apart from its predecessors. Central to Mamba's design are its Selective-State-Spaces (SSM): these are recurrent models that selectively process information based on the current input, effectively filtering out irrelevant data to focus on what is most critical for efficient processing. Additionally, Mamba simplifies its architecture by replacing the complex attention and MLP blocks in Transformers with a single, unified SSM block, enhancing inference speed and reducing computational load. Mamba incorporates hardware-aware parallelism, using a specially designed parallel algorithm that optimizes recurrent operations for improved hardware efficiency, potentially boosting performance even further.


Figure \ref{fig:variants} shows the three variants of our Mamba-based model that we experiment with. The first (MACE-V) is a vanilla Mamba model with two vertical Mamba layers. The second (MACE-H) is a Mamba model with two concatenated/horizontal models. The third is our proposed model, BMACE, which presents the structure of our bidirectional Mamba network.  Bidirectional Mamba blocks and fully-connected layers are the main modules in the network. It processes a 10-second audio signal as a Constant Q Transform (CQT) feature. The model integrates a fully-connected layer into the input, which then proceeds to two Mamba blocks with opposite masking directions, represented as dotted boxes in Figure \ref{fig:variants}. The outputs from these blocks are concatenated and passed through a fully-connected layer to maintain the input's original dimensions. We added residual operation in the blocks and layers to increase the information entropy.

\section{Experiment}\label{sec:experiment}

\subsection{Experiment Setting}
Our models are implemented with Pytorch\cite{paszke2019pytorch} framework. All experiments are conducted on the instance node at Lambda\footnote{\url{https://cloud.lambdalabs.com/instances}} that has a single NVIDIA RTX A6000 GPU (24 GB), 14vCPUs, 46 GiB RAM and 512 GiB SSD. 
Our model was trained and validated on the MARL annotations\footnote{\url{https://github.com/tmc323/Chord-Annotations}}
of uspop2002 dataset\cite{berenzweig2004large}. Each 10-second audio signal was processed with a 5-second overlap between consecutive signals. The signals were sampled at 22,050 Hz and analyzed using a Constant Q Transform (CQT) that covered 6 octaves starting from C1, with 24 bins per octave and a hop size of 2048. The CQT features were then converted to log amplitude using the formula $S_{\text{log}} = \ln(S + \epsilon)$ , where 
S represents the CQT feature, and $\epsilon$
is an extremely small number. This was followed by the application of global z-normalization, using the mean and variance derived from the training data. 

We evaluate the three versions of our model described in Section \ref{sec:mamba} (MACE-V, MACE-H, and BMACE) against state of the art CRNN-based \cite{McFeeB17_StructTrainingChords_ISMIR} and transformer-based \cite{park_bi-directional_2019} models on the 25-label maj-min and the 170-label large chord vocabularies.


\subsection{Results}
Table \ref{tab:uspop} presents the model validation results. BMACE performs slightly better on some label types than the CRNN and BTC models, though the difference is not likely to be statistically significant. However, there is a notable improvement over the non-bidirectional Mamba models (MACE-V and MACE-H). The most significant differences are in the size and processing requirements of the various models, as shown in Table \ref{tab:uspop}. As previously observed, CRNNs are more efficient than transformer-based models, and the claim that Mamba networks are also more efficient holds true. All Mamba-based models use only 1/25th of the parameters of the transformer-based BTC model and are slightly less than 1/3 smaller than CRNN. This reduction in the number of parameters is reflected in the lower GFlops required to run the model.

\newpage
\section{Acknowledgments}
This work is supported by the National Science Foundation (NSF) grant 2228910. The opinions expressed in this work are solely those of the authors and do not necessarily reflect the views of the NSF.
\bibliography{ISMIRtemplate}

%
%
%
%
%

\end{document}